\RequirePackage{ifpdf}
\ifpdf 
\documentclass[pdftex]{sigma}
\else
\documentclass{sigma}
\fi

\numberwithin{equation}{section}

\def\dd{\partial}
\def\half{\frac{1}{2}}
\def\dd{\partial}
\def\a{\alpha}
\def\b{\beta}
\def\c{\gamma}
\def\d{\delta}

\newcommand{\p}{^{\prime}}

\begin{document}

\allowdisplaybreaks

\renewcommand{\thefootnote}{$\star$}

\renewcommand{\PaperNumber}{029}

\FirstPageHeading

\ShortArticleName{Jordan--Schwinger Representations and Factorised Yang--Baxter Operators}

\ArticleName{Jordan--Schwinger Representations\\
and Factorised Yang--Baxter Operators\footnote{This paper is a contribution to the Proceedings of
the XVIIIth International Colloquium on Integrable Systems and Quantum
Symmetries (June 18--20, 2009, Prague, Czech Republic).  The full
collection is
available at
\href{http://www.emis.de/journals/SIGMA/ISQS2009.html}{http://www.emis.de/journals/SIGMA/ISQS2009.html}}}

\Author{David KARAKHANYAN~$^\dag$ and Roland KIRSCHNER~$^\ddag$}

\AuthorNameForHeading{D.~Karakhanyan and R.~Kirschner}

\Address{$^\dag$~Yerevan Physics Institute,
Br. Alikhanian Str.~2, 375036 Yerevan, Armenia}
\EmailD{\href{mailto:karakhan@lx2.yerphi.am}{karakhan@lx2.yerphi.am}}

\Address{$^\ddag$~Institut f\"ur Theoretische
Physik, Universit\"at Leipzig,
PF 100 920, D-04009 Leipzig, Germany}
\EmailD{\href{mailto:Roland.Kirschner@itp.uni-leipzig.de}{Roland.Kirschner@itp.uni-leipzig.de}}

\ArticleDates{Received October 28, 2009, in f\/inal form March 30, 2010;  Published online April 07, 2010}

\Abstract{The construction elements of the factorised form of the
Yang--Baxter $R$ operator acting on generic representations of
$q$-deformed $s\ell(n+1)$ are studied. We rely on the
iterative construction of such representations by the restricted
class of Jordan--Schwinger representations.
The latter are formulated explicitly. On this basis
the parameter exchange and intertwining operators
are derived.}


\Classification{81R50; 82B23}

\section{Introduction}

Generic representations of $s\ell_q (n+1)$ can be
iteratively constructed from Jordan--Schwinger ones of
$s\ell_q (m+1)$, $m=1, \dots ,n$,  \cite{BL} as an application
of the  method of induced representations \cite{GN,BW}.
This iterative relation is conveniently formulated in
terms of Lax matrices \cite{Derkachov:2008fe},
the Yang--Baxter solutions intertwining
a generic with the fundamental representation. The generic representation
of $g\ell (n+1)$ may be labelled by $\ell_1, \dots , \ell_{n+1}$,  the
eigenvalues of the Cartan subalgebra elements in action on the lowest weight
states.
 $\sum \ell_i$ refers to the $U(1)$ factor. The Lax matrix
will be denoted by $L(u_1, \dots, u_{n+1})$, where the spectral parameter
$u$ is combined with the representation parameters to $u_i = u + \ell_i$.

 The Yang--Baxter operator $R$ acting on the tensor product
of two generic representations is def\/ined by the Yang--Baxter relation
with two Lax matrices,
\begin{gather}
\label{YB}
 R L(u_1, \dots , u_{n+1}) L(v_1, \dots , v_{n+1}) = L(v_1, \dots , v_{n+1})
L(u_1, \dots , u_{n+1}) R.
\end{gather}
The solution of the Yang--Baxter relation has to be
treated in the framework of the quantum group theory, the emergence
of which was initiated by solving problems of quantum integrable
systems. The $q$-deformed algebra and co-algebra relations are encoded
in the particular Yang--Baxter relation of the fundamental
$R$ matrix with Lax matrices (for a comprehensive review see e.g.~\cite{Isaev}).
In \cite{KhT} the universal $R$ operator has been constructed,
covering the case $s\ell_q (n+1)$ in particular, in algebraic terms
based on the combinatorial structure of root systems.
In \cite{DJM, BK} the solutions of Yang--Baxter relation  have been
constructed and studied with restriction to cyclic representations of
$s\ell_q (n+1)$ and with the algebra relations modif\/ied for
application to the generalised chiral Potts model.
In \cite{DGZ} a method for constructing the $R$ operator in
spectral decomposition has been developed applicable to
tensor products of af\/f\/inisable representations of (super) algebras.

The generic Yang--Baxter operator (\ref{YB})
can be constructed as a product of  operators
permuting pairs of parameters.
As elementary factor operators
one may use the ones permuting adjacent parameters of one Lax matrix,
e.g.~$W_i(u_i,u_{i+1})$ permuting $u_i$ with $u_{i+1}$,  and the
operator $F$ permuting $u_{n+1}$ with $v_1$.
The former are the intertwiners
transforming between equivalent representations
with representation labels appearing in dif\/ferent order.
The latter intertwine between equivalent tensor product
representations.

In  studies of the chiral Potts model \cite{BS,HY,T1,BOU}
concerning cyclic representations
of $s\ell_q(2)$ the $R$ operator is represented as a product of
four Boltzmann weights which are identif\/ied as intertwiners of
equivalent cyclic representations.  In the extension to
the generalised chiral Potts model involving cyclic representations
of $s\ell_q (n+1)$ this factorisation is used~\cite{DJM,BK,Kashaev,T2}.

Motivated by applications of integrable chains to problems of
gauge f\/ield theory \cite{L,BDM}, where inf\/inite-dimensional representations
appear, we consider realisations of representation generators
in terms of Heisenberg canonical pairs. In the $s\ell(2)$ case
one pair, $x$, $\dd$, $[\dd, x] =1 $ is suf\/f\/icient and  representations
are spanned by monomials $x^m$ with $1$ representing the lowest weight
vector. The factorisation of the $R$ operator can be observed also in this
formulation. For example, its integral kernel appears factorised into
four symmetric two-point functions being the kernels of intertwining
operators \cite{KKM}. The  construction of the  $R$
operator with $s\ell (2)$ symmetry in
terms of canonical pairs and by factors permuting representation
parameters in the product of Lax matrices has been proposed in~\cite{Derkachov:2005fg}, applied to the study of Baxter operators in~\cite{Derkachov:2005be}
 and worked out for the rational, trigonometric and elliptic cases in~\cite{Derkachov:2007gr}.

In the extension of this approach to higher rank the class of
Jordan--Schwinger representations plays the key role.  This notion stands
for the realisation of the $s\ell(n+1)$ algebra generators in terms of
$n+1$ canonical pairs described below and representation modules spanned
by monomials of $x_i$ with $1$ representing the lowest weight vector.
Generic representations can be constructed iteratively based on the ones of
Jordan--Schwinger type.
We consider the trigonometric case with generic values of
the  deformation parameter $q$, excluding roots of 1. Then these modules
do not involve cyclic representations and appear just as deformations
of the classical representation modules~\cite{GN}.

A treatment of the rational ($q=1$) case is given in
\cite{Derkachov:2006fw}; in the present paper
the part of the results concerning
the intertwining and exchange operators
is extended to the quantum case.
The case $s\ell_q(3)$ has been studied  in~\cite{DKKV}
relying on direct calculations instead of the detailed Lax matrix
factorisation considered here.

In this paper we obtain the permutation and intertwining operators
being the construction elements for Yang--Baxter $R$ operators
by the factorisation outlined above.
The iterative construction
of the generic Lax matrix from Jordan--Schwinger ones helps to
f\/ind the intertwi\-ners~$W_i$, because the calculation  reduces to the
analysis of the lower rank $g\ell(i+1)$ case.
Also the parameter permutation  operator $F$ can be obtained by solving
the reduced problem,
where the Lax matrices are substituted by the Jordan--Schwinger ones in two
forms. Indeed one can do the iterative construction in one way such that the
Jordan--Schwinger Lax matrix involving~$v_1$ appears as the left-most factor or
in the other way such that $u_{n+1}$ appears as the right-most factor.
Then the permutation of $u_{n+1}$ with $v_1$ is determined by these
Jordan--Schwinger Lax matrices only.

In the next section we formulate the Jordan--Schwinger representations
\cite{J,S} of
$s\ell_q(n+1)$ and the corresponding Lax matrix in several versions.
In Section~\ref{section3} we derive factorised forms  of the Jordan--Schwinger
Lax matrix. The dependence on the parameters $u_+ = u + 2\ell$,
$u_- = u-1 $ appears separated, one enters by a left matrix factor and the
other by a right one. Therefore the
factorised forms of the Lax matrix are useful for the simplif\/ication
of the relation def\/ining the permutation operator $F$.
The def\/ining relation for
$F$ and its solution are discussed in detail in Section~\ref{section4}. The def\/ining
relations can be represented as the discrete and deformed analogon of
a wave
equation for a chain of $n+1$ sites. The solution $F$ is
analogous to a  standing wave, the multiplicative superposition of a
forward and a backward travelling wave expressed in terms of the
$q$-exponential.
Further, in Section~\ref{section5},
the intertwining operators $W_i$ are obtained. We f\/ind f\/irst
operators $D_i$ changing the Lax matrix with $\ell_i$, $\ell_{i+1}$ to
the one with $\ell_i +1$, $\ell_{i+1} - 1 $ up to some remainders,
and this leads to $W_i$ as a power of $D_i$.

\section[Jordan-Schwinger representations]{Jordan--Schwinger representations}\label{section2}

With $n+1$ Heisenberg conjugated pairs we can construct
operators generating the $q$-deformed $gl(n+1)$ algebra.
These generators
induce by lowest weight construction
representations
of  the restricted Jordan--Schwinger type,
where a constant function represents the
vector of lowest weight.
We start at the def\/initions of the operators
\[
 E_{ij}^J = \frac{x_i}{x_j} [N_j],\qquad
E_{ij}^{-J} = - \frac{x_j}{x_i} [N_j],\qquad
E_{ij}^{TJ} = - [N_i]\frac{x_j}{x_i},\qquad
E_{ij}^{-TJ} =  [N_i]\frac{x_i}{x_j},
\]
for $i \not = j$, $i,j = 1, \dots , n+1$.
$N_i = x_i \dd_i$ acts as inf\/initesimal
dilatation operator on the coordinate operator $x_i$.
The square bracket denotes the $q$-number of the entry,
$ [N] = \frac{q^N - q^{-N}}{q - q^{-1}}$.

These def\/initions result in 4 versions (distinguished by the label
$J$, $-J$, $TJ$, $-TJ $) of the Jordan--Schwinger representation of $g\ell_q (n+1)$.
The operators with superscripts $-J$, $-TS$ are related to the
corresponding ones with $J$, $TJ$ by inversion of coordinates
$x_i \to x_i^{-1}$ with the
corresponding transformation of the derivative operators.
The operators with superscript $TJ$, $-TJ$ are related to the
corresponding ones $J$, $-J$ by transposition def\/ined like
result of partial integration, $x_i^T = x_i$, $N_i^T = -N_i -1$,
and a shift of $N_i$
proportional to the identity operator, $N_i + 1 \to N_i$.

The commutation relation between the operators of
the same superscript are close to the $g\ell_q(n+1)$
algebra relations,
$ [E_{ij}^J,  E_{jk}^J]_{q^{\pm1}} =
q^{\mp N_j} E_{ik}^J
$ for $i \not = k$ and
$ [E_{ij}^J,  E_{ji}^J]_{1} = [N_i - N_j] $.
Here we use an appropriate modif\/ication of the
commutator notation
def\/ined  as $[A, B]_{q} = A B - q B A $.
The relations for operators with superscript $-TJ$ are the
same and the ones for the operators with the superscript
$TJ$ or $-J$ are obtained by substituting
the factor on r.h.s.\ of the f\/irst relation as
$ q^{\mp N_j}\rightarrow q^{\pm N_j} $
and supplementing the r.h.s.\ of the second
relation by a minus sign.
These operators can be related to the Chevalley basis
of the $ sl_q(n+1) $ algebra as
\[
e^C_i = E_{i,i+1}^C, \qquad f_i^C = E_{i+1,i}^C, \qquad 2 h_i^C =
(-1)^{|C|} [N_i - N_{i+1}],
\quad i= 1, \dots , n.
\]
$C$ stands for $J$, $-J$, $TJ$, $-TJ$ and in the last relation
the sign is dependent on the superscript as $(-1)^{|C|}$
where $|C| = 0$ for $C=J, -TJ$ and $|C| = 1$ for
$C= -J, TJ$.
The algebra relations including Serre's relations can
be checked.
Alternatively one can extend the construction to the
Cartan--Weyl generators by def\/ining them by $q$-commutators
iteratively,
\begin{gather}
\label{iter}
 E_{i,i+1} = e_i, \qquad E_{i+1,i} = f_i
\\
E_{ij} = [E_{i, j-1}, E_{j-1,j} ]_{q},\quad i+1<j, \qquad
E_{ij} = [E_{i, i-1}, E_{i-1,j} ]_{q^{-1}}, \quad i>j+1.\nonumber
\end{gather}
In the cases $ X= J, -JT$ this leads to
\begin{gather*}
 E_{ij}^{JS} = q^{-(N_{i+1} + \dots + N_{j-1})} E_{ij}^J,  \quad i<j,
\\
E_{ij}^{JS} = q^{(N_{i-1} + \dots + N_{j+1})} E_{ij}^J, \quad i>j,\quad
\quad i,j = 1, \dots , n+1.\nonumber
\end{gather*}
The same relations apply to the case $-JT$. The resulting
generators are distinguished by adding to the superscript
the letter $S$. The relation for the cases $JT$, $-J$ are
obtained by modifying the factors
on r.h.s.\ by changing the signs in the exponents of~$q$.
We def\/ine also
\[ E_{ii}^C = (-1)^{|C|} N_i. \]
 We study the Lax matrix in Jimbo's form~\cite{Jimbo:1985zk} where the
matrix elements are related to the generators as
\begin{gather*}
 L_{ij} (u) = q^{-(u-\half) - \half (E_{ii} + E_{jj})}
E_{j,i} , \quad i> j,
\\
 L_{ij} (u) = q^{+(u-\half) + \half (E_{ii} + E_{jj})}
E_{j,i}, \quad i< j, \qquad
 L_{ii} (u) = [ u + E_{ii} ].
 \end{gather*}
Substituting the Cartan--Weyl generators constructed in four
versions $CS= JS$, $ -JS$, $JTS $, $-JTS$ we obtain
\begin{gather}
\label{LaxJS}
L^{CS}_{ij} (u) = q^
{(u-\half) (-1)^{|C|} N_{ij})}
E_{j,i}^C ,  \quad i< j,
\\
L^{CS}_{ij} (u) = q^
{-(u-\half) (-1)^{|C|} N_{ji})}
E_{j,i}^C , \quad  i > j, \qquad
L^{CS}_{ii} (u) = [ u + (-1)^{|C|} N_i ].
\end{gather}
We use the notation $N_{ij} $ def\/ined as
 \[ N_{ij} = \tfrac 12 N_i + N_{i+1} +\dots + N_{j-1} +
\tfrac 12 N_j, \]
where the addition on the indices is evaluated
$\mod(n+1)$.
We shall use also
\[ N_{ij} + N_{ji} = \sum_1^{n+1} N_s = 2 \hat \ell, \qquad
\tfrac 12 (N_{ij} - N_{ji} ) = N_{i,j}\p. \]
These versions of Jordan--Schwinger Lax matrices
allow  factorised forms where the coordinates and the
dilatation operators are separated. Dif\/ferent forms
give preference to some index value $i = 1, \dots , n+1$.
In the case $JS$ we introduce the set
of quantum coordinates
\begin{gather}
\label{Xq}
 X_i = q^{-N_{i,n+1}} x_i, \quad i= 1,\dots , n+1, \qquad N_{n+1,n+1} = 0.
\end{gather}
Then we obtain
\begin{gather}
\label{DLDJS}
L^{JS}(u) = \hat X^{-1} \tilde L^{JS} (u) \hat X, \qquad
\hat X = {\rm diag}\, (X_1, \dots, X_{n+1}).
\end{gather}
The central factor involves the dilatation operators
$N_i$ only
\begin{gather}
\label{cJS}
\tilde L^{JS} (u)  =
 \left(
 \begin{array}{ccccc}
[u-1 + N_1]   &q^{u-1} [N_1] & \dots &
q^{u-1} [N_1] &  q^{u-1} [N_1]  \\
q^{1-u} [N_2]
 &[u-1 +N_2 ] &  \dots &
q^{u-1} [N_2] &  q^{u-1} [N_2]  \\
\dots &\dots &\dots &\dots &\dots  \\
q^{1-u} [N_{n+1}] & q^{1-u} [N_{n+1}]& \dots &
q^{1-u} [N_{n+1}] & [u-1 + N_{n+1}]
\end{array} \right ).
\end{gather}
In the case $-JS$ we have with the same def\/inition
(\ref{Xq})
\begin{gather}
\label{DLD-JS}
L^{-JS}(u) = \hat X \tilde L^{-JS} (u) \hat X^{-1},
\end{gather}
where $\tilde L^{-JS} (u)$ is obtained from
$\tilde L^{JS} (u)$ by substituting $N_i$ by $-N_i$.

In the cases $TJS$ and $-TJS$ we introduce the set
of quantum coordinates
\[
X_i^T = x_i q^{-N_{i,1}}, \quad i= 1, \dots , n+1.
\]
These coordinates can be factorised in terms of  diagonal matrices
similar to (\ref{DLDJS}), (\ref{DLD-JS}),
\begin{gather*}
L^{TJS}(u) = \hat X_T \tilde L^{TJS} (u)
\hat X_T^{-1}, \qquad
L^{-TJS}(u) = \hat X_T^{-1} \tilde L^{-TJS} (u)
\hat X^T,
\\
\hat X_T = {\rm diag}\, (X^T_1, \dots, X^T_{n+1}).\nonumber
\end{gather*}
The central factor is in the case $-TJS$
\[
\tilde L^{-TJS}  (u) =
\left(
 \begin{array}{ccccc}
[u-1 + N_1]   &q^{u-1} [N_2] & \dots &
q^{u-1} [N_n] &\!\! q^{u-1} [N_{n+1}]  \\
q^{1-u} [N_1]
 &[u-1 +N_2 ] &  \dots &
q^{u-1} [N_n] &\!\! q^{u-1} [N_{n+1}]  \\
\dots &\dots &\dots &\dots &\dots  \\
q^{1-u} [N_{1}] & q^{1-u} [N_{2}]& \dots &
q^{1-u} [N_{n}] & [u-1 + N_{n+1}]
\end{array} \right )
\]
and the corresponding matrix for the case $TJS$ is
obtained from the latter by changing the sign in front of
all~$N_i$.

\section{Factorisation of Lax matrices}\label{section3}

The simple form of the Jordan--Schwinger representation Lax matrix
(\ref{LaxJS}) leads to factorised expressions. Let us formulate and proof
the details in the version $JS$.
We introduce the following notations
\begin{gather*}
\Lambda = q^{2 \hat N\p_{*,n+1}}, \qquad
\hat N\p_{*,n+1} = {\rm diag}\, (N_{1, n+1}\p, \dots ,
N_{n, n+1}\p, -\hat \ell),\nonumber\\
 \hat B = q^{\hat N}  \Lambda, \qquad
\hat N = {\rm diag}\, (N_1, \dots , N_{n+1}),
\nonumber\\
M_{1,-1} (A) = \hat m_{1,-1} - A \sigma_-.
\end{gather*}
We use the standard matrices $\sigma_- = \hat e_{n+1,1}$, $\hat m_1 $
and $\hat m_{1,-1}$,
\begin{gather}
\label{m1-1}
\hat m_{1} =
 \left(
 \begin{array}{ccccc}
1  &0 & \dots &
0 &\!\! 0  \\
0
 &1 &  \dots &
0 &\!\! 0  \\
\dots &\dots &\dots &\dots &\dots  \\
0 &0 & \dots & 1 & 0   \\
1 &1 & \dots & 1 & 1   \\
\end{array}\!\!\! \right ), \qquad
\hat m_{1,-1} =
 \left(
 \begin{array}{ccccc}
1  &-1 & \dots &
0 &\!\! 0  \\
0
 &1 &  \dots &
0 &\!\! 0  \\
\dots &\dots &\dots &\dots &\dots  \\
0 &0 & \dots & 1 & -1   \\
0 &0 & \dots & 0 & 1   \\
\end{array}  \right )
\end{gather}
and its inverse $\hat m_{1,-1}^{-1} $ with all upper-triangular elements
including the
diagonal equal to $1$ and the other equal to $0$.

\begin{proposition}
The Jordan--Schwinger form of the Jordan--Schwinger Lax matrix \eqref{LaxJS}
can be written as
\begin{gather}
\label{LJSM}
L^{JS} (u) = \hat X^{-1} \Lambda^{-1}
M_{1,-1} \big(q^{-2(u-1+ 2\hat \ell )}\big)
  \hat B
[u-1] M_{1,-1}^{-1} (q^{u-1}) \hat X.
\end{gather}
\end{proposition}

\begin{proof}
We start the proof from the triangular factorisation used earlier
\cite{Derkachov:2008fe}.
The central Lax matrix factor (\ref{cJS})  can be represented as
a product of 3 matrices, the central factor being
upper triangular and the f\/irst and third factors
being lower triangular of special type
\begin{gather*}
\tilde L^{JS} (u)  = M_L (u) K^{JS} (u) M_R (u),
\\
M_R (u) = D_R^{-1} (u-1) \hat m_1 D_R (u-1), \qquad
M_L (u) = D_L^{-1} \left ( M_R(u) \right )^{-1}
 D_L,
\nonumber\\
D_R (u) = {\rm diag}\, (q^{-2u}, \dots , q^{-2u}, 1),\qquad
D_L = {\rm diag}\, ( q^{\alpha_1}, \dots , q^{\alpha_n}, 1), \qquad
 \alpha_i = - 2 N_{n+1,i}.\nonumber
\end{gather*}
The matrix $\hat m_1$ has unit elements on the diagonal and
on the last row with all others vanishing.
The upper-triangular central factor is given by
\begin{gather}
 K^{JS} (u)  =
 \left(
 \begin{array}{ccccc}
[u-1] q^{N_1}  &  \lambda [u-1] [N_1] & \dots &
\lambda [u-1] [N_1] &  q^{u-1} [N_1]  \\
0
 &[u-1] q^{N_2} &  \dots &
\lambda [u-1] [N_2] &  q^{u-1} [N_2]  \\
\dots &\dots &\dots &\dots &\dots  \\
0 &0 & \dots & 0 & k^*   \\
\end{array}  \right ),\nonumber\\
\label{KJS}
 k^* = q^{N_{n+1}- \sum_1^{n+1} N_s}
\left[u-1 + \sum_1^{n+1} N_s\right].
\end{gather}
The essential ingredients of the f\/irst
and the third factors of the triangular factorisation
enter
in terms of diagonal matrices $D_R (u)$, $D_L$.
  The lower-triangular form
is provided by the standard matrix~$\hat m_1$~(\ref{m1-1}).

Now we observe that the central upper-triangular factor can
be further factorised and the same diagonal feature appears.
The form of $K^{JS}(u) $ (\ref{KJS}) suggests
the f\/irst step of further factorisation
\begin{gather}
\label{KJSf}
K^{JS} (u) =
{\rm diag}\big(q^{N_1}, \dots , q^{N_n}, k^*\lambda q^{1-u} \big)
\tilde K^{JS}
\,{\rm diag}\left([u-1], \dots ,[u-1], \frac{q^{u-1}}{\lambda} \right).
\end{gather}
Then we use the standard matrix $\hat m_{1,-1}$,
 to obtain
\begin{gather}
\label{tildeKJSf}
\tilde K^{JS} = B^{-1}  \hat m_{1, -1}  B
\hat m_{1,-1}^{-1}
B  = {\rm diag}\, (\dots q^{\beta_i} \dots ),
\end{gather}
$\beta_i $ are determined up to a constant,
$
\beta_i = 2\sum_i^{n+1} N_s + c$, $i=1,\dots , n+1$.
We choose $ c= -N_{n+1} - \sum_1^{n+1} N_s $, thus
\[ \beta_i = N_i + 2 N_{i,n+1}\p, \quad i=1, \dots , n, \qquad
\beta_{n+1} = - \sum_1^n N_s.
\]
We consider now the complete factorised expression for
the Lax matrix, $L(u) = \hat X^{-1} M_L K M_R \hat X$.
It is appropriate to modify the def\/inition of the
left and right lower-triangular factors
including one of the diagonal factors into which
$K^{JS}$ has been decomposed now
\[
M_R\p (u) = {\rm diag}\left(1, \dots , \frac{q^u}{\lambda [u]}\right)
M_R (u+1).
\]
We obtain
\[
L^{JS} (u) = \hat X^{-1} \Lambda^{-1}
M_R^{\prime -1} (u-1+ 2\hat \ell )
\hat m_{1,-1}  B  \hat m_{1,-1}^{-1}
[u-1] M_R\p (u-1) \hat X .
\]
The two steps (\ref{KJSf}) and (\ref{tildeKJSf}) result in the
factorisation of the upper-triangular central fac\-tor~$K^{JS} (u)$
where the dependence on the operators $N_i$ and on the
spectral parameter $u$ enter in terms of
diagonal matrices.

The lower triangular special matrix
$\hat m_1 $
can be related to the upper triangular
special mat\-rix~$\hat m_{1,-1}$
with the help of $\sigma_- = \hat e_{n+1,1}$
\[
M_R^{\prime -1} (u) = M_R^{-1} (u+1) \,{\rm diag}\big(1,\dots ,1, 1-q^{-2u} \big)
= M_{1,-1} \big(q^{-2u}\big) \hat m_{1,-1}^{-1}.
\]
This allows to write
the Jordan--Schwinger form of the Lax matrix  as has been claimed,
completing the proof.
\end{proof}

The representation constraint
\begin{gather}
\label{repc}
\Phi_c = \sum_1^{n+1} N_s = (-1)^{|C|} 2 \ell \hat I
\end{gather}
reduces the algebra to a simple one and f\/ixes the representation
of $s\ell_q(n+1) $ which is irreducible for generic values of $\ell$.
The constraint operator commutes with $N_i$ but not with the coordinate
operators. Actually the Lax matrices depend only on coordinate ratios
which commute with~$\Phi_c$. Therefore we transform the obtained
factorisation formulae f\/irst into forms expressed in terms
of ratios of coordinates dividing out one of them.
Then the constraint can be imposed simply by substituting the
related $N$ operator by the expression
obtained from the constraint equation~(\ref{repc}).
In the case $JS$ we write
\begin{gather}
 X_i = x_{n+1} q^{\gamma} X_i\p,\nonumber\\
\label{DXp}
 X_i\p = q^{-\hat \ell - \half -
N_{i,n+1}\p -  \gamma}
\frac{x_i}{x_{n+1}},\qquad
\tilde X_i\p = q^{\hat \ell + \half +
N_{i,n+1}\p - \tilde \gamma}
\frac{x_i}{x_{n+1}},\quad i= 1, \dots , n ,
\end{gather}
and modify the def\/inition of the diagonal
matrices of the coordinates as
\[
\hat X = {\rm diag}\, (X_1\p,\dots ,X_n\p,1),\qquad
\hat X^d = {\rm diag}\, (\tilde X_1\p,\dots ,\tilde X_n\p,1),\qquad
\Gamma (\gamma) = {\rm diag}\, (1, \dots ,1, q^{-\gamma}).
\]
Now $x_{n+1}$ can be cancelled in the expression for the Lax matrix
(\ref{LJSM}) and the representation constraint can be imposed by simply
replacing the operator $\hat \ell$ by the number $\ell$.
The form of the spectral parameter dependence
becomes more symmetric
after the shift $ u\p = u+ \ell$
and it is convenient
to introduce $u_+ = u\p + \ell$, $u_- = u\p -1-\ell$.
Then we have
\begin{gather}
\label{LJSpm}
L^{JS}(u_+,u_-) = \hat X^{d -1}
\Gamma^{-1} (\tilde \gamma)
 M_{1,-1} \big(q^{-2u_+}\big)
  \hat B  M_{1,-1}^{-1}\big( q^{-2u_- }\big)
\Gamma(\gamma)  \hat X.
\end{gather}
In the case $TJS$ we write
\begin{gather}
\label{DX-Tp}
X_{Tj}\p = q^{\hat \ell - N_{i 1+}\p -\gamma_T}
\frac{x_j}{x_1}, \qquad
\tilde X_{Tj}\p = q^{-\hat \ell + N_{i 1+}\p -
\tilde \gamma_T}
\frac{x_j}{x_1}, \quad j= 2, \dots , n+1
\\
\hat X_T = {\rm diag}\, (1, X_{T2}\p, \dots , X_{T n+1}),
\qquad
\hat X^d_T = {\rm diag}\, (1, \tilde X_{T2}\p, \dots ,
\tilde X_{T n+1}),
\nonumber\\
\Gamma_T (\gamma) = {\rm diag}\, (q^{-\gamma}, 1, \dots , 1),\qquad
\hat B_T = {\rm diag}\big( q^{-\sum_2^{n+1} N_s},
q^{N_2 - 2N_{2,1}\p},\dots ,q^{N_{n+1} - 2N_{n+1,1}\p} \big).
\nonumber
\end{gather}
Now $x_1$ can be cancelled in the expression for the Lax matrix
and the representation constraint can be imposed by
eliminating $N_1$ in favour of $\hat \ell$ and
replacing the operator $\hat \ell$ by the number~$-\ell$
\begin{gather}
\label{LTJSpm}
L^{TJS}(u_-,u_+) = \hat X_{T}\Gamma_T(\gamma)
  M_{1,-1}^{-1} \big(q^{-2u_-}\big)
  \hat B_T^{-1}  M_{1,-1} \big(q^{-2u_+ }\big)
\Gamma^{-1}_T(\tilde \gamma_T)  \hat X_{T}^{d -1}.
\end{gather}

\section{Parameter exchange operator}\label{section4}

\subsection{The def\/ining relation}

We consider the def\/ining equation for an operator
interchanging representation parameters in the product
of two Jordan--Schwinger Lax matrices,
\begin{gather}
\label{Finter}
\hat F L_y (u_-, u_+)  L_x (v_+, v_-)
 =  L_y (u_-, v_+)  L_x (u_+, v_-) \hat F.
\end{gather}
Let us substitute $L_y (u_-, u_+)$ by the
Lax matrix in the version $TJS$ (\ref{LTJSpm}),
$L_y (u_-, u_+) = L^{TJS} (u_-,u_+)$
with the def\/initions given in (\ref{DX-Tp})
and the substitution of the canonical pairs by
$y_i$, $\dd_{y i}$ as indicated by subscript~$y$
and  $L_x (v_+, v_-)$
by the one in the version $JS$ (\ref{LJSpm}),
$L_x (v_+, v_-) = L^{JS} (v_+,v_-)$ with the
def\/initions given in (\ref{DXp}).
We rely on the fact that in all cases the quantum
coordinates $X_i\p$ commute with all coordinate
operators $\tilde X_j\p$ and  try as ansatz
\[
\hat F = F\big(\hat Y_T^d,  \hat X^{d}\big).
\]
The def\/ining condition reduces to
\begin{gather*}
\hat F \hat B_{Ty}^{-1} M_{1,-1} \big(q^{-2 u_+}\big)
\Gamma_T^{ -1} (\tilde \gamma_T)
\hat Y_T^{d  -1}
\hat X^{d  -1}  \Gamma^{ -1}(\tilde \gamma) M_{1,-1} \big(q^{-2v_+}\big)  \hat B_x\\
\qquad {}=
\hat B_{Ty}^{-1} M_{1,-1} \big(q^{-2 v_+}\big)
\Gamma_T^{ -1} (\tilde \gamma_T)
\hat Y_T^{d  -1}
\hat X^{d -1}  \Gamma^{ -1} (\tilde \gamma)
M_{1,-1} \big(q^{-2u_+}\big)  \hat B_x \hat F.
\end{gather*}
The condition is now specif\/ied by the assumption that
the phases  depend on
the spectral and the representation parameters as
$ \tilde \gamma(v_+)$, $\tilde \gamma_T (u_+)$,
on l.h.s.\ and that also the arguments $u_+$, $v_+$
in the phases are subject to the
interchanging operation by~$\hat F$.

\begin{proposition}
The parameter exchange operator $\hat F$ acting as in \eqref{Finter}
on the product of Jordan--Schwinger Lax matrices of version $TJS$
\eqref{LTJSpm} for $L_y$ and $JS$ \eqref{LJSpm} for $L_x$ and with
specifying $\tilde \gamma(u) = \tilde \gamma_T(u) = u$ has the form
\begin{gather*}
F= (X_1^d )^{u_+-v_+} \ e_{q^2} ( q^{u_+-v_++1} Z)
 e_{q^{-2}} (- q^{v_+ - u_+ -1} Z),
\\
Z= \sum_2^n q^{v_+-1} \tilde X_j\p \tilde Y_{Tj}\p
X_1^{\prime -1 } +
\tilde  Y_{T n+1}\p \tilde X_1^{\prime -1},\nonumber
\end{gather*}
$e_{q^2} (Z)$ denotes the deformed exponential.
The definition of $\tilde X\p_j$, $\tilde Y\p_{T j}$ has been given in
\eqref{DXp}, \eqref{DX-Tp}.
\end{proposition}

The proof will be arranged in three parts. In the f\/irst part,
up to the end of this subsection, the def\/ining conditions are written in a
convenient form. The result is reminicent of a chain with $n+1$
sites where the boundaries at sites 1 and $n+1$
give extra contributions as compared to the
bulk region. In the bulk, outside the boundary sites, we f\/ind
 $q$-deformed discrete waves. This is the second step of the proof
given in Subsection~\ref{section4.2}. In the f\/inal step, in Subsection~\ref{section4.3},
we show how the boundary conditions f\/ix the
wave number and result in a multiplicative superposition
of travelling waves.

The def\/ining condition can be transformed to
\begin{gather*}
\hat Y_T^{d } \Gamma_T(v_+) M^{-1}_{1,-1} \big(q^{-2 v_+}\big)
\hat B_{Ty}\hat F \hat B_{Ty}^{-1}
M_{1,-1} \big(q^{-2 u_+}\big) \hat \Gamma_T^{ -1} (u_+)
Y_T^{d  -1}
\\
\qquad{}=
\hat X^{d  -1}  \Gamma^{ -1}(u_+) M_{1,-1} \big(q^{-2u_+}\big)
\hat B_x  \hat F  \hat B^{-1}_x
M^{-1}_{1,-1} \big(q^{-2v_+}\big) \Gamma (v_+)
\hat X^{d }.\nonumber
\end{gather*}
The quantum coordinates $X_i\p$, $Y_{T i}\p$ do not enter
the resulting  condition. We shall suppress  temporarily primes
and the signs
on the dual coordinates $\tilde X_i\p$, $\tilde Y_{Ti}\p
\hat X^d$, $\hat Y_T^d$ indicating the deviation from
$X_i\p$, $Y_{T i}\p$.
We consider the transformation of the arguments of
$\hat F$
by
$\hat B_{Ty} \hat F \hat B^{-1}_{Ty} $,
\begin{gather*}
q^{\beta_1^T} \hat Y_T q^{-\beta_1^T}
=\hat Y_T \hat b_{T 1} =
\hat Y_T \,{\rm  diag}\big(1,q^{-1},\dots , q^{-1}\big), \\
q^{\beta_i^T} \hat Y_T q^{-\beta_i^T}
= \hat Y_T   \hat b_{T i}=
\hat Y_T \,
{\rm diag}  \big(1,q,\dots ,q , \stackrel{i} q,\dots , q^{-1}\big),
\quad i= 2, \dots , n+1.
\end{gather*}
The transformation by $\hat B_x  \hat F  \hat B^{-1}_x $
acts similar,
\begin{gather*}
q^{\beta_i} \hat X q^{-\beta_i}
= \hat X   \hat b_i = \hat X \,{\rm  diag}  \big(q^{-1},\dots ,q^{-1} , \stackrel{i} q,
\dots , q, 1\big), \quad i= 1, \dots , n, \\
q^{\beta_{n+1}} \hat X q^{-\beta_{n+1}}
= \hat X   \hat b_{n+1} =
\hat X \, {\rm diag}\big(q^{-1},\dots , q^{-1},1\big).
\end{gather*}
The commutation relation of the quantum coordinates
are
\[ X_i X_j = q X_j X_i, \quad
1 \le i < j \le n,
\qquad
 Y_{Ti} Y_{T j} = q Y_{Tj} Y_{Ti}, \quad
2 \le i < j \le n+1.
\]
This results in an analogous pattern for
the similarity transformations:
\begin{gather*}
 X_i \hat X X_i^{ -1} =
\hat X   \hat s_i =\hat X \, {\rm diag} \big(q^{-1}, \dots ,\stackrel{i} 1, q, \dots q,1\big),
\quad i= 1, \dots , n,
\\ Y_{Ti} \hat Y_T Y_{Ti}^{  -1} =
\hat Y_T   \hat s_{T i} =
\hat Y_T \, {\rm diag}\big(1, q^{-1}, \dots ,\stackrel{i} 1, q, \dots q\big),
\quad i= 2, \dots , n+1.
\end{gather*}
We complete the def\/initions by $\hat s_{n+1} = \hat I$,
$\hat s_{T 1} = \hat I $.
Then the action on $\hat F$ is
\[
\hat B_{Ty} \hat F \hat B^{-1}_{Ty} = {\rm diag} \big(\dots F^y_i \dots \big),
\qquad \hat B_x  \hat F  \hat B^{-1}_x =
{\rm diag}\big(\dots F^x_i \dots \big),
\]
where the following abbreviations are convenient
\begin{gather*}
 F^y_i = F\big(\hat X, \hat Y \hat b_{T i} \big), \qquad
F^x_i = F\big(\hat X \hat b_{ i}, \hat Y \big),
\\
 Y_{T j} F^y_i Y^{ -1}_{T j} =
F\big(\hat X, \hat Y \hat b_{T i} \hat s_{T j} \big)
= F^y_{i,j},\qquad
X_j^{ -1} F^x_i X_{j} =
F\big(\hat X \hat b_{ i} \hat s^{-1}_j , \hat Y \big)
= F^x_{i,j}.
\end{gather*}
The abbreviations allow to write the def\/ining
condition as
\begin{gather}
\label{defineFG}
 \Gamma_T (v_+) \{ {\rm l.h.s.} \} \Gamma_T^{ -1} (u_+)
=
\Gamma^{ -1} (u_+) \{ {\rm r.h.s.} \} \Gamma (v_+),
\\
\{ {\rm l.h.s.} \}
= {\rm diag} \big (F^y_{j-1, j}\big) + \hat Y_T \hat m_{1,-1}^{-1}
\hat Y_T^{ -1} \,{\rm diag}\big( F^y_{j,j} - F^y_{j-1, j}\big)\nonumber\\
{}+ \frac{q^{-2v_+}}{ 1-q^{-2 v_+} }
\big(  \hat Y_T \hat M_1 \hat Y_T^{ -1}
\,{\rm diag} \big( F^y_{j,j} - F^y_{j-1, j}\big)
+ \big(1- q^{-2 (u_+ - v_+)} \big) \hat Y_T
\hat m_{1,-1}^{-1}
\hat \sigma_- \hat Y_T^{ -1} F^y_{n+1,1}
\big),\nonumber\\
\{ {\rm r.h.s.} \}
= {\rm diag} \big(F^x_{j+1, j}\big) + {\rm diag} \big( F^x_{j,j} -
F^x_{j+1, j}\big) \hat X^{ -1} \hat m_{1,-1}^{-1}
\hat X
\nonumber\\
{}+ \frac{q^{-2v_+}}{ 1-q^{-2 v_+} }
\big(  {\rm diag} \big( F^x_{j,j} -
F^x_{j+1, j}\big) \hat X^{ -1} \hat M_1
\hat X
+ \big(1- q^{-2 (u_+ - v_+)} \big) F^x_{1,n+1}
\hat X^{ -1} \sigma_- \hat m_{1,-1}^{-1} \hat X
\big).\nonumber
\end{gather}
We have used the properties of the matrices
$M_{1,-1} (q^{-2u})$ in particular the
decomposition of their products
with diagonal matrices.

\subsection{Travelling waves}\label{section4.2}

 Let us consider f\/irst the equations
resulting  from the matrix elements $(i,j)$ with
$i, j \not = 1, n+1$.
From the diagonal we have
\begin{gather}
\label{diagj}
 F^y_{j,j} - q^{-2v_+} F^y_{j-1,j} =
F^x_{j,j} - q^{-2v_+} F^x_{j+1,j}.
\end{gather}
Since
\begin{gather}
\label{bsj}
 \hat b_{T j} \hat s_{T j} = \hat b_j \hat s_j^{-1}
= {\rm diag}\big(1, \dots , 1, \stackrel{j} q, 1,\dots ,1\big) ,
\\
\hat b_{T j-1} \hat s_{T j} = \hat b_{j+1}
\hat s_j^{-1}
= {\rm diag}\big(1, \dots , 1, \stackrel{j} {q^{-1}}, 1,\dots ,1\big),
\quad j= 2, \dots , n, \nonumber
\end{gather}
these equations are fulf\/illed by requiring the
equality of the
f\/irst terms on both sides and of the second terms
\begin{gather}
\label{diagF}
 F^y_{j,j}  = F^x_{j,j}, \qquad
  F^y_{j-1,j} =  F^x_{j+1,j}
\end{gather}
and this implies
$ F= F(\hat X \cdot \hat Y) $, i.e.\ the dependence
assumed in the ansatz is specif\/ied as on the $n+1$
products of the diagonal elements,
  $ X_1, X_2 Y_{T2}, \dots , X_n Y_{T n}, Y_{T n+1} $.

The non-diagonal bulk matrix elements in (\ref{defineFG})result in
the conditions
\begin{gather}
\label{ndiagj}
 Y^{-1}_{T j} \big( F^y_{j,j} - F^y_{j-1,j} \big) X_j^{-1}
=
Y^{-1}_{T i} \big( F^x_{i,i} - F^x_{i+1,i} \big) X_i^{-1},\quad
i,j \not = 1, n+1.
\end{gather}
We have obtained a multiplicatively formulated
dif\/ference equation close to a wave equation.
Instead of a displacement of the position variable at
site $i$ we have its multiplication by $q^{\pm 1}$,
and we f\/ind  solutions corresponding to
waves of such multiplicative displacements
going forward or backward in the chain.

In order to f\/ind a particular solution we try
the ansatz
\[ F_+ = \prod_{k \uparrow} f (Z_k), \qquad Z_k =  X_k Y_{T k}.
 \]
Then l.h.s.\ of (\ref{ndiagj}) results in
\[ Y_{T j}^{-1} \big(F^y_{j,j}- F^y_{j-1,j}\big) X^{-1}_j
=  \prod^{j-1} f(q Z_k)
\left (f(q Z_j) - f \big(q^{-1} Z_j\big) \right ) Z^{-1}_j
  \prod_{j+1} f(q Z_k),
\]
and this becomes independent of $j$ provided
\begin{gather}
\label{f+}
\big(f(q Z_j) - f \big(q^{-1} Z_j\big)  \big) Z^{-1}_j
= c_+\p f(q Z_j).
\end{gather}
This can be rewritten as
\[ f(q^{-2} Z) = \left(1- \frac{c_+\p}{q} Z\right)   f(Z),
\]
i.e.\ the functional equation for the
$q$-deformed exponential. We set $c_+\p=q c_+$ and obtain
\[
 f_+(Z) = e_{q^{-2}} (c_+ Z), \qquad
 F_+ = e_{q^{-2}} \left(c_+ \sum X_j Y_j\right).
\]
The multiplication property of the $q$-exponential~\cite{qexp},
 $e_q(V) e_q(U) = e_q (U+V)$ for
$U V = q V U$  allows to write the products
in $F_{+}$ in terms of a $q$-exponential again.
The alternative ansatz
\[ F_- = \prod_{k \downarrow} f (Z_k), \qquad
\ Z_k =  X_k Y_{T k} \]
leads to the condition
\begin{gather}
\label{f-}
\big(f(q Z_j) - f \big(q^{-1} Z_j\big)  \big) Z^{-1}_j
= c_-\p f\big(q^{-1} Z_j\big)
\end{gather}
for the independence of $j$.
We obtain (for $c_-\p= -q^{-1} c_-$)
\[
 f_-(Z) = e_{q^2} (c_- Z), \qquad
 F_- = e_{q^{2}} \left(c_- \sum X_j Y_j\right).
\]
The boundary conditions result in a
combination of these solutions. The multiplicative
form of the def\/ining equations results in a
product instead of a sum.
Before analysing the boundary we show that indeed the product
of the obtained solutions of (\ref{ndiagj}) is a solution too.
This is true if the chain has one site only,
\begin{gather*}
 Z^{-1} \big( f_+ (q Z) f_- (q Z) - f_+ \big(q^{-1} Z\big) f_- \big(q^{-1} Z\big) \big)
\\
\qquad{}= Z^{-1} \big( f_+ (qZ) - f_+ \big(q^{-1} Z\big) \big) f_- (qZ) + f_+\big(q^{-1} Z\big)
Z \big(f_- (qZ) - f_- \big(q^{-1} Z\big) \big) \\
\qquad{}=
c_+\p f_+ (q Z) f_- (qZ) + c_-\p f_+
\big(q^{-1} Z\big) f_-\big(q^{-1} Z\big) =
 ( c_+ + c_-) f_+(qZ) f_- \big(q^{-1} Z\big).
\end{gather*}
In the last step the functional equation of the $q$-exponentials
and  $c_{\pm}\p = \pm q^{\pm 1} c_{\pm}$
has been used. The proof for the general case
of the chain with many sites can be done
relying on the multiplication property of the
$q$-exponential.

\subsection{Boundary conditions}\label{section4.3}

We reconsider the equations arising from the diagonal
now focussing on the boundary. For $i=j=n+1$ and
$i=j=1$ we have from (\ref{defineFG})
\begin{gather}
\label{diag1}
F^y_{n+1,n+1} - q^{-2v_+} F^y_{n,n+1} =
q^{\gamma^d (u_+) - \gamma^d (v_+) }
\big( F^x_{n+1,n+1} - q^{-2 u_+} F^x_{1,n+1}\big),\\
q^{\gamma_T^d (u_+) - \gamma_T^d (v_+) }
\big( F^y_{1,1} - q^{-2 u_+} F^y_{n+1,1}\big)
= F^x_{1,1} - q^{-2v_+} F^x_{2,1}.\nonumber
\end{gather}
The relations (\ref{bsj}) extend to the boundary as
\begin{gather*}
 \hat b_{T 1} \hat s_{T 1} = q^{-1} \hat b_1
\hat s^{-1}_1, \qquad
\hat b_{T n+1} \hat s_{T n+1} = q \hat b_{n+1}
\hat s^{-1}_{n+1},
\\ \hat b_{T n+1} \hat s_{T 1} = q \hat b_2
\hat s^{-1}_1, \qquad
\hat b_{T n} \hat s_{T n+1} = q^{-1} \hat b_{1}
\hat s^{-1}_{n+1}.\nonumber
\end{gather*}
A consistent solution is possible if we specify the
dependence of the additional phases as
\[ \tilde \gamma (u) = \tilde \gamma_T (u) = u. \]
Then the  boundary diagonal equations are fulf\/illed if
\begin{gather}
\label{diag1F}
F_{1,1}^y q^{u_+ - v_+} = F_{1,1}^x, \qquad F_{n+1,1}^y q^{v_+ - u_+} =
F_{2,1}^x, \\
 F_{n+1,n+1}^y  = q^{u_+ - v_+} F_{n+1,n+1}^x, \qquad F_{n,n+1}^y  =
q^{v_+ - u_+}   F_{1,n+1}^x.\nonumber
\end{gather}
These suf\/f\/icient conditions are solved by
\begin{gather}
\label{FZX1}
 F= X_1^{u_+ - v_+} F\p (Z_2, \dots ,Z_n, Z_{n+1}),
\\
Z_k = X_k Y_k X_1^{-1} , \quad k= 2,\dots ,n, \qquad Z_{n+1} = q^{\alpha} Y_{n+1} X_1^{-1}.\nonumber
\end{gather}
This is compatible with the bulk solution to the
non-diagonal equations because $c_{\pm}$ may depend on the
variables referring to the boundary.

We can write the non-diagonal equations $(i,j)$ in (\ref{defineFG})
in the form
\begin{gather*}
 Y_{T j}^{  -1} q^{\delta_{j, 1} u_+}
\big( F^y_{j,j} - q^{-\delta_{j,1} 2 (u_+ - v_+)} F^y_{j-1,j}
\big)  q^{\delta_{j,n+1} v_+ } X_j^{ -1}
\\
\qquad{}=
Y_{T i}^{  -1} q^{\delta_{i, 1} v_+}
\big( F^x_{i,i} - q^{-\delta_{i,n+1} 2 (u_+ - v_+)} F^x_{i+1,i}
\big)  q^{\delta_{i,n+1} u_+ } X_i^{ -1}  .
\end{gather*}
The diagonal equations have been solved by relating
$F^y_{j,j}$, $F^y_{j-1,j}$ to $F^x_{j,j}$, $F^x_{j+1,j}$
as in~(\ref{diagF}), (\ref{diag1F}).
This reduces the non-diagonal conditions to
\begin{gather}
\nonumber
Y_{T j}^{-1} \big (F^y_{j,j} - F^y_{j-1,j} \big)
X_j^{-1} =
q^{u_+} \big(F^y_{1,1} - q^{-2 (u_+ - v_+)} F^y_{n+1,1} \big)
X_1^{-1}  \\
\qquad{}=
Y_{T n+1}^{-1} \big(F^y_{n+1,n+1} - F^y_{n,n+1} \big)
q^{v_+}.\label{ndiag1}
\end{gather}
We start now with the ansatz compatible with the
bulk conditions (\ref{diagj}), (\ref{ndiagj})
and with the diagonal boundary conditions~(\ref{diag1})
\begin{gather}
\label{F+F-}
 F = X_1^{u_+ -v_+} F_+ (Z\p) F_-(Z\p), \qquad
F_{\pm} (Z\p) = e_{q^{\mp 2}} (c_{\pm} Z\p), \qquad
Z\p= \sum_2^{n+1} Z_k.
\end{gather}
The f\/irst factor takes care of the diagonal boundary relations but is
not relevant in the following. We shall see that the product form with
specif\/ic constants $c_{\pm}$ results only from the relation involving $j=1$.
First we check that the specif\/ication
of the argument $Z_k$ in (\ref{FZX1})
does not invaliditate the
condition of independence of $j$ of the
f\/irst expression in (\ref{ndiag1})
\[
Y_{T j}^{-1} \big(F_{\pm j,j}^y - F_{\pm j-1,j}^y
\big) X_j^{-1} =
c_{\pm}\p F_{\pm} \big(q^{\pm 1} Z\big) X_1^{-1} q. \]
The relation involving $j=n+1$ is obeyed by $F_{\pm}$ separately and
also by the product; it f\/ixes just $\alpha$ in (\ref{FZX1})
because the dif\/ference
equation for the factor $f_{\pm} ( Z_{n+1})$ is modif\/ied to
\[ q^{v_+ + \alpha} Z^{-1} \big(f_{\pm} (q  Z) - f_{\pm} \big(q^{-1} Z\big) \big) =
c_{\pm}\p q f_{\pm} \big(q^{\pm 1} Z\big) \]
and for $\alpha = -v_+ + 1$ the dif\/ference to the
corresponding equations
for $j= 2,\dots , n$ (\ref{f+}), (\ref{f-}) disappears.

Notice that $b_{T1} s_{T1} = {\rm diag} (1,q^{-1} , \dots , q^{-1} )$ and
$b_{T,n+1} s_{T 1} = {\rm diag} (1, q, \dots , q)$.
 The condition involving $j=1$ is
\begin{gather*}
 q^{v_+}  \big( q^{u_+-v_+} F_+ \big(q^{-1} Z\p\big)
F_-\big(q^{-1} Z\p\big) -
q^{v_+-u_+} F_+ (q Z\p) F_-(Z\p) \big) \\
\qquad{} =
( c_+\p +c_-\p) q F_+ (q Z\p) F_-\big(q^{-1} Z\p\big).
\end{gather*}
It is solved by f\/ixing $c_{\pm}\p$ in (\ref{F+F-})
by the following conditions
\[
 q^{u_+ - v_+ } c_+\p - q^{v_+ -u_+ } c_-\p = 0, \qquad
 \lambda [u-v] q^{v_+} = (c_+\p + c_-\p) q .
\]
This leads to
the result for the operator interchanging the
parameters $u_+$, $v_+$ in the product of two
Jordan--Schwinger Lax matrices as
formulated in the proposition
 $(Z= q^{v_+ -1} Z\p)$, completing the proof.

\section{Intertwiners}\label{section5}

The formulation of the iterative construction
of a generic Lax matrix
from Jordan--Schwinger ones takes to introduce canonical pairs
$x_{i,k}$,  $\dd_{i,k}$, $i= 1, \dots , k+1$, $k=1, \dots ,n$.
The corresponding  generators  and Lax matrices
will be denoted by superscript $k$, being constructed
from the canonical pairs $x_{i,k_1}$, $\dd_{i,k_1}$,
$k_1 \le k $ as above.

The Lax matrix representing $ g\ell_q (n+1)$
is obtained in terms of the one of $ g\ell_q (n)$
as
 \begin{gather}
\label{Ln+1}
[u-1] L^{(n)}(u)= L^{JS,n}(u)L^{n-1 \prime}(u-1),
 \end{gather}
where the f\/irst factor is the $ g\ell_q (n+1)$ Jordan--Schwinger Lax
matrix $(n+1) \times (n+1)$ and
the second factor is calculated in terms of the
Lax matrix $L^{(n-1)}$ of $ g\ell_q (n)$
as
\begin{gather}
\label{Lprime}
L^{n-1 \prime}(u)= L^{(n-1)}_{\a\b}(v)  \hat e_{\a\b}
+q^{-u}\mathcal {A}^{(n-1)}_\a    \hat e_{n+1, \a}
+[u] \hat e_{n+1 n+1}, \quad \a, \b, \c = 1, \dots , n,
\\
\mathcal{A}^{(n-1)}_\a=-\sum_{\c=1}^{n} L^{(n-1)}_{\c\a}(0) X_{\c,n}.\nonumber
\end{gather}
At each iteration step one representation parameter
enters by imposing the representation constraint (\ref{repc}).
This involves to change the coordinates to their ratio
$x_{i,k} \to x_{i,k}\p = \frac{x_{i,k}}{x_{k+1,k}} $ and
to eliminate $N_{k+1}^{(k)}$ in favour of $\sum_1^{k+1} N_s^{(k)} =
\ell_{k+1}$.
We use the notation $N_i^{(k)} = x_{i,k} \dd_{i,k}$.
The parameter $\ell_{n+1}$ does not enter the second
factor in~(\ref{Ln+1}).

In this way the result depends besides of the
spectral parameter on the representation parameters
$\ell_1, \dots , \ell_{n+1}$ and
we write $L^{(n}) (\ell_1, \dots , \ell_{n+1}|u)$.

We shall construct  intertwiners $W_m$, $m=1, \dots ,n$ acting as
\begin{gather}
\label{WmLn}
 W_m L^{(n)} (\ell_1, \dots ,\ell_m,\ell_{m+1}, \dots  \ell_{n+1}|u) =
 L^{(n)} (\ell_1, \dots ,\ell_{m+1}, \ell_m,\dots  \ell_{n+1}|u) W_m.
\end{gather}
For $n=1$ the complete Lax operator is obtained from (\ref{Ln+1}), (\ref{Lprime})
with the trivial $1\times 1$ matrix
$L^{(0)} (\ell_1|u) = [u+\ell_1]$
 \begin{gather*}
L^{(1)}(\ell_1,\ell_2|u)=\left(\begin{array}{cc}[u+E^{(1)}_{11}] &q^{u+\frac12(E^{(1)}_{11}+
E^{(1)}_{22}-1)}E^{(1)}_{21}\vspace{1mm}\\ q^{-u-\frac12(E^{(1)}_{11}+ E^{(1)}_{22}-1)}
E^{(1)}_{12}&[u+E^{(1)}_{22}] \end{array}\right).
\end{gather*}
We have
\[  D_1=E^{(1)}_{21}= \frac1{x_{1,1}}
\big[N_{1}^{(1)}\big], \qquad
N^{(1)}_{1}\equiv x_{1,1}\dd_{1,1}
\]
and  the remaining $g\ell_q(2)$ generators are given by
 \[
 E^{(1)}_{11}=\ell_1
+N^{(1)}_{1}, \qquad E^{(1)}_{1 2}=x_{1,1}\big[\ell_2-\ell_1-N^{(1)}_{1}\big],
 \qquad E^{(1)}_{22}=\ell_2-N^{(1)}_{1}. \]
Then we check easily
 \begin{gather}\label{int12}
D_1 L^{(1)}_{\a\b}(\ell_1,\ell_2|u)=L^{(1)}_{\a\b}(\ell_1+1,\ell_2-1|u)D_1+ [\ell
_2-\ell_1] \d^2_\a\d^1_\b q^{-u-\frac12(\ell_1+\ell_2-1)},
 \end{gather}
 and further
 \begin{gather*}
D_1^n L^{(1)}_{\a\b}(\ell_1,\ell_2|u)=L^{(1)}_{\a\b}(\ell_1+n,\ell_2-n|u)D_1^n\nonumber\\
\phantom{D_1^n L^{(1)}_{\a\b}(\ell_1,\ell_2|u)=}{}
+[n] [\ell_2-\ell_1+1-n] \d^2_\a\d^1_\b
  q^{-u-\frac12\big(E^{(1)}_{11}+ E^{(1)}_{22}-1\big)}, 
 \end{gather*}
from which one deduces that
\begin{gather*}
 W_1 = D_1^{\ell_2-\ell_1+1}.
\end{gather*}
Passing to the $g\ell_q(3)$ Lax matrix $L^{(2)}$ according to
(\ref{Ln+1}) one sees that the same operator
$W_1 =D_1^{\ell_2-\ell_1+1}$
intertwines the representation parameters $\ell_1$, $\ell_2$ there.
Taking into account
that the f\/irst factor in~(\ref{Ln+1}) commutes with $D_1$ and
does not depend on $\ell_1$ and $\ell_2$, while the second factor obeys~(\ref{int12}) one concludes that (\ref{int12}) becomes:
 \begin{gather*} 
D_1L^{(2)}_{\a\b}(\ell_1,\ell_2,\ell_3|u)=
L^{(2)}_{\a\b}(\ell_1+1,\ell_2-1,\ell_3|u)D_1
\\
\qquad{}
+[\ell_2-\ell_1]_q\Big(q^{-u-\frac12(E^{(2)}_{11}+E^{(2)}_{22}-1
)}\d^2_\a\d ^1_\b-q^{-u-\frac12(E^{(2)}_{11}+E^{(2)}_{33}-1
)}\d^3_\a\d^1_\b x_{23}q^{E^{(2)}_{33}-E^{(1)}_{22}}\Big),\nonumber
\end{gather*}
here $E^{(2)}_{\a\b}$ stand for the $g\ell_q(3)$ generators and the
intertwining operator is again given by~$D_1^{\ell_2-\ell_1+1}\!$.
The
analogous relation in $g\ell_q(4)$ case looks like:
 \begin{gather*}
D_1L^{(3)}_{\a\b}(\ell_1,\ell_2,\ell_3,\ell_4|u)=
L^{(3)}_{\a\b}(\ell_1+1,\ell_2-1,\ell _3,\ell_4|u)D_1+
\\
\qquad{}+[\ell_2-\ell_1]_q\Big(q^{-u-\frac12(
E^{(3)}_{11}+ E^{(3)}_{22}-1)}\d^2_\a\d^1_\b\nonumber\\
\qquad{}-q^{
-u-\frac12(E^{(3)}_{11}+E^{(3)}_{33}-1)}\d^3_\a\d^1_\b
\Big( x_{2,2}q^{E^{(2)}_{33}-
E^{(1)}_{22}}+(q^2-1)\frac{x_{2,3}}{x_{3,3}}
\big[N^{(3)}_{3}\big]\Big) \nonumber\\
\qquad{}-q^{-u-\frac12(E^{(3)}_{11}+E^{(3)}_{44}-1)}\d^4_\a\d^1_\b\Big(x_{2,3}
q^{E^{(3)}_{44}-
E^{(2)}_{22}-N^{(3)}_3 + 1 }-x_{2,2}x_{3,3}q^{E^{(3)}_{44}-
E^{(1)}_{22}+1}\Big)\Big),\nonumber
\end{gather*}
$E^{(3)}_{\a\b}$ stand for generators of $g\ell_q(4)$.

This  means that
by  recurrent construction  the $g\ell_q(n+1)$ Lax
matrix  inherits the $\ell_1$ and $\ell_2$ intertwining operator
form the $g\ell_q(2)$ case for any $n>2$.
Indeed, because  Jordan--Schwinger Lax matrix commutes with
$D_1$  while the second matrix multiplier in (\ref{Ln+1})
produces inhomogeneous
terms upon commutation with $D_1$ only in the f\/irst column.
$D_1$ does not commute only with~$\mathcal{A}_1^{(n)}$. Then the
matrix multiplication rule tells that inhomogeneous terms can appear
only in f\/irst column of complete Lax operator and this proves the
assertion.

We pass now to the operator $D_2$ intertwining $\ell_2$ and $\ell_3$
 \begin{gather*}
D_2=\frac1{x_{2,2}}\big[N^{(2)}_{2}\big] q^{N^{(2)}_{1}-N^{(1)}_{1}}+
\frac{x_{1,1}}{x_{1,2}}\big[N^{(2)}_{1}\big].
 \end{gather*}
One f\/inds that the wanted homogeneous transformation rule:
 \begin{gather}\label{int23}
D_2L_{\a\b}(\ell_1,\ell_2,\ell_3|u)=L_{\a\b}(\ell_1,\ell_2+1,\ell_3
-1|u)D_2,
 \end{gather}
in $g\ell_q(3)$ case is violated in two matrix elements of the Lax
operator, $L_{32}$ and $L_{31}$, by appearance in
the r.h.s.\ of (\ref{int23}) of inhomogeneous terms
\[
[\ell_3-\ell_2] q^{_u-\frac12(E^{(2)}_{22}+E^{(2)}_{33}-1)}
\qquad
{\rm and}
\qquad
x_{1,1}[\ell_3-\ell_2]q^{_u-\frac12(E^{(2)}_{22}+E^{(2)}_{33}+
N^{(2)}_{1}-N^{(2)}_{2}-1)},
\]
correspondingly.

The next intertwining operator is calculated from
 \begin{gather*}
D_3=q^{N^{(3)}_{1}+N^{(3)}_{2}-N^{(2)}_{1}-N^{(2)}_{2}}\frac1{x_{3,3
}}\big[N^{(3)}_{3}\big]+q^{N^{(3)}_{1}-N^{(2)}_{1}}\frac{x_{2,2}}{x_{2,3}}
\big[N^{(3)}_{2}\big]+\frac{x_{1,2}}{x_{1,3}}\big[N^{(3)}_{1}\big].
 \end{gather*}
The considered examples lead us to the
\begin{proposition}\label{proposition3}
The intertwining operator $W_m$ def\/ined by \eqref{WmLn} can be constructed as
\begin{gather*}
W_m=D_m^{\ell_{m+1}-\ell_m+1},
\end{gather*}
where
\begin{gather}
D_{m}=\frac{x_{1,m-1}}{x_{1,m}}\big[N^{(m)}_{1}\big]+
q^{N^{(m)}_{1}-N^{(m-1)}_{1}}
\frac{x_{2,m-1}}{x_{2,m}}\big[N^{(m)}_{2}\big]\nonumber\\
\phantom{D_{m}=}{} +
q^{N^{(m)}_{1}-N^{(m-1)}_{1}+N^{(m)}_{2}-N^{(m-1)}_{2}}
\frac{x_{3,m-1}}{x_{3,m}}\big[N^{(m)}_{3}\big]+\cdots\nonumber
\\
\phantom{D_{m}=}{}
 + q^{\sum\limits_{k=1}^{m-2}(N^{(m)}_{k}-N^{(m-1)}_{k})}
\frac{x_{m-1,m-1}}{x_{m-1,m}}\big[N^{(m)}_{m-1}\big]
+
q^{\sum\limits_{k=1}^{m-1}(N^{(m)}_{k}-N^{(m-1)}_{k})}
\frac{1}{x_{m,m}} \big[N^{(m)}_m\big].\label{Dm}
\end{gather}
\end{proposition}

\begin{proof}
We start the proof with the case $m=n$ and check f\/irst the intertwining
relation in the undeformed case $q \to 1$, where
 \begin{gather*}
D_{n}|_{q=1}=\dd_{n,n}+x_{n-1,n-1}
\dd_{n-1,n}+\dots+x_{1,n-1}\dd_{1n}.
 \end{gather*}
The ansatz for $q\not = 1$ is obtained by substituting
the derivatives as
$\dd_{i,n} \to x_{i,n}^{-1} [N_i^{(n)}] $ and including a
factor $q^{\alpha_i}$  in each term. $\alpha_i$ are then
determined to result in (\ref{Dm}) by demanding that
\begin{gather*}
D_{n}L^{(n)}_{\a\b}(\ell_1,\ell_2,\ldots,\ell_{n},\ell_{n+1}|u)=
L^{(n)}_{\a\b}(
\ell_1,\ell_2,\ldots,\ell_{n}+1,\ell_{n+1}-1|u)D_{n} + \cdots ,
 \end{gather*}
where the dots stands for the contribution to
 the single matrix element $L_{n+1,n}$,
for which the
r.h.s.\ contains the additional inhomogeneous term
$[\ell_{n+1}-\ell_n]
q^{-u-\frac12(E_{nn}^{(n)}+E_{n+1n+1}^{(n)}-1)}$.
This remainder disappears in the corresponding relation for
$W_n = D_n^{\ell_{n+1}-\ell_n +1}$. Note that the canonical pairs
with $k<n-1$ are passive in this step.

In the case $m<n$ similar remainders appear in the commutation
relation of $D_m$ with $L^{(n)} $ in the matrix elements
$L_{i j }$,  $1\leq j\leq m$, $m+1\leq i\leq n+1 $. It is suf\/f\/icient to
consider $i=m+1$, $j=m $ and check the vanishing of these remainders
for the commutation with $W_m$.
The commutation with other matrix elements does
not result in more conditions because the
corresponding generators $E_{ij}$,
$1\leq i\leq k$, $k+1\leq j\leq n+1 $ are obtained from
$E_{k+1,k}$ by (\ref{iter}).
In this way $W_m$ interchanges the representation labels
$\ell_m$, $\ell_{m+1} $ not only in $L^{(m)} $ but in all
Lax matrices  $L^{(n)}$, $n\ge m $,  proving the assertion.
\end{proof}

\section{Discussion}\label{section6}

The intertwining operators $W_i$ and the parameter exchange operator
$F$ can be used to obtain the generic Yang--Baxter $R$ operator in
a factorised form. This provides a convenient
approach because the def\/ining conditions for these factors are
much simpler compared to the one for $R$.

It is instructive to compare with the treatment of the
rational ($q=1$) case \cite{Derkachov:2006fw} in detail,
although the reduction to Jordan--Schwinger
representations was not used there.
In both cases the Lax matrices are factorised to
separate coordinate from shift operators.
In our case we are lead to modif\/ied coordinate
operators with $q$-deformed commutation relations.
The intertwiners are given by powers of operators,
where the powers are determined from the
representation labels in the same way. How the
operators appearing in the $q=1$ case are deformed is
clearly seen in Proposition~\ref{proposition3}.
In the rational case the exchange operator appears
as a power calculated from representation labels
of a coordinate dif\/ference expression.
The expression involving the $q$-exponential is the
appropriate deformation for the restricted
case of Jordan--Schwinger representations.

The representation spaces can be spanned on polynomial functions of
$x_{i,k}\p$, on which the above canonical pairs of operators act by
multiplication or dif\/ferentiation, the constant function
representing the lowest weight vector of a generic representation.
This polynomial representation works for the algebra elements and
also for the complete $R$ operator. However the action of the~$W_i$,~$F$ as constructed above lead  from polynomials
to branched functions, i.e.\ to quite dif\/ferent realisations of the algebra
representations. This may be regarded as less important as long as
the latter appear merely as construction elements of $R$, but
requires more investigation in general.

We observe similarities in the result for $F$ and $W_i$. Both
are expressed in terms of the set canonical pairs of two Jordan--Schwinger
representations. It should be possible to relate both
expressions by transformations of these canonical pairs.

We have obtained  construction elements needed for the treatment of
integrable systems def\/ined by a generalised Heisenberg
spin chain with the ordinary spin replaced by
generic representations of $s\ell_q (n+1)$.

\subsection*{Acknowledgements}

We are grateful to S.E. Derkachov and P. Valinevich
for joint work and useful discussions.
The work has been supported in part by
Deutsche Forschungsgemeinschaft (KI 623/6-1).

\pdfbookmark[1]{References}{ref}
\LastPageEnding

\end{document}